\documentclass[jcp,reprint]{revtex4-1}

\usepackage{graphicx}
\usepackage{dcolumn}
\usepackage{bm}
\usepackage{amsmath}
\usepackage{amssymb}

\usepackage[utf8]{inputenc}
\usepackage[T1]{fontenc}
\usepackage{mathptmx}
\usepackage{etoolbox}
\usepackage{float}
\makeatletter
\def\@email#1#2{%
 \endgroup
 \patchcmd{\titleblock@produce}
  {\frontmatter@RRAPformat}
  {\frontmatter@RRAPformat{\produce@RRAP{*#1\href{mailto:#2}{#2}}}\frontmatter@RRAPformat}
  {}{}
}%
\makeatother
\begin{document}

\renewcommand{\vec}[1]{{\bf{#1}}}

\title[]{Assessing models of force-dependent unbinding rates via infrequent metadynamics}

\author{Willmor J. Pe\~{n}a Ccoa}
\author{Glen M. Hocky}%
 \email{hockyg@nyu.edu}
\affiliation{ 
Department of Chemistry, New York University, New York, NY 10003
}%
\date{\today}
\begin{abstract}
Protein-ligand interactions are crucial for a wide range of physiological processes. 
Many cellular functions result in these non-covalent `bonds' being mechanically strained, and this can be integral to proper cellular function.
Broadly, two classes of force dependence have been observed---slip bonds, where unbinding rate increases, and catch bonds where unbinding rate decreases.
Despite much theoretical work, we cannot we predict for which protein-ligand pairs, pulling coordinates, and forces a particular rate dependence will appear.
Here, we assess the ability of MD simulations combined with enhanced sampling techniques to probe the force dependence of unbinding rates.
We show that the infrequent metadynamics technique correctly produces both catch and slip bonding kinetics for model potentials.
We then apply it to the well-studied case of a buckyball in a hydrophobic cavity, which appears to exhibit an ideal slip bond.
Finally, we compute the force-dependent unbinding rate of biotin-streptavidin.
Here, the complex nature of the unbinding process causes the infrequent metadynamics method to begin to break down due to the presence of unbinding intermediates, despite use of a previously optimized sampling coordinate.
Allowing for this limitation, a combination of kinetic and free energy computations predict an overall slip bond for larger forces consistent with prior experimental results, although there are substantial deviations at small forces that require further investigation.  
This work demonstrates the promise of predicting force-dependent unbinding rates using enhanced sampling MD techniques, while also revealing the methodological barriers that must be overcome to tackle more complex targets in the future.   
\end{abstract}
\maketitle
\section{Introduction}
\label{sec:intro}
Mechanical forces play an important role in a wide range of biological processes \cite{Bell618,SOKURENKO2008314,thomas2008catch,geiger2009environmental,oakes2014stressing,persat2015mechanical,murrell2015forcing,makarov2016perspective,zimmermann2017mechanoregulated,cox2018bacterial,zimmermann2019feeling,freedman2019mechanical,gomez2021molecular}. 
Cells have evolved \textit{mechanosensing} mechanisms by which the behavior of a protein or protein complex changes in a stereotypical way in response to that applied force. 
In general, these forces produce two types of results: they can have a \textit{thermodynamic} effect on the conformational landscape of the protein(s) or a \textit{kinetic} effect, changing reaction rates \cite{hartmann2020infinite,gomez2021molecular}.
In this work, we will focus on the kinetic effects of force on protein-ligand unbinding \cite{Bell618,thomasSokurenko,makarov2016perspective}.  
Although much work has been done experimentally and theoretically to understand the role of mechanosensitive unbinding rates in biological processes \cite{thomas2008catch,thomasSokurenko,manibog2014resolving,huang2017vinculin,oleg,barsegov2005dynamics,dudko2008theory,makarov2016perspective,chakrabarti2017phenomenological,novikova2021evolving}, much less is known about the molecular details that contribute to the force dependence of the rate.
Here, we wish to assess whether molecular dynamics (MD) simulations coupled with enhanced sampling techniques are suitable for this task.

Protein-ligand interactions are essential in mediating cellular adhesion and cell-cell interactions. 
These non-covalent ``bonds'' are put under tension due to the action of molecular motors in the cellular cytoskeleton and/or mediated by tension in the cellular membrane \cite{Bell618,grashoff2010measuring,oakes2014stressing,lenne2021cell}. 
Crucially, at short time scales we can think of these forces as quasi-static, with forces typically in the piconewton scale for each bond. 
Although MD has been used to probe the effect of force on proteins or even protein-ligand interactions \cite{best2001can,isralewitz2001steered,okimoto2017evaluation,stirnemann2013elasticity,languin2018three,Rico6594}, to our knowledge it has not been used to predict equilibrium unbinding kinetics under these quasistatic, small force conditions.
As recently reviewed, this regime is particularly challenging because these small forces are not expected to substantially shift the behavior of the system outside the linear response regime, hence sampling has to be very accurate to capture the subtle structural changes leading to large changes in rate \cite{gomez2021molecular}. 
Theoretical work and coarse grained studies in this area have typically focused on (free) energy surfaces representing the unbound and possibly multiple bound states of the system, given the difficulty of probing these systems at a fully molecular level \cite{oleg,dudko2011locating,makarov2016perspective,chakrabarti2017phenomenological}.

The biggest challenge to predicting bond lifetimes is that the relevant time scales for dissociation may be on the order of milliseconds to tens of seconds for systems that we are interested in, meaning that we would not expect to see any unbinding events within a standard MD simulation \cite{gomez2021molecular}. 
We were inspired by a large amount of recent literature on the development of enhanced sampling MD techniques designed to predict the unbinding time of drug molecules from their protein targets \cite{dicksonun,tiwary2017and,wang2018frequency,pramanik2019can,ray2020kinetics,ahn2020ranking}. 
These techniques accelerate the unbinding of the ligand by many orders of magnitude in such a way that many unbinding events can be observed within the limitations of standard computational resources, and allow for statistical reweighting of the observed unbinding times to predict their unbiased values.
Approaches to generating rare unbinding events can be broadly broken into two categories, (1) those that simulate many copies of the system and select only trajectories that advance along some progress variable, and (2) those that push the ligand out of its binding pose by applying an energy bias in the bound state.

Here we report our results from using Infrequent Metadynamics (InfrMetaD), a method that computes unbinding times from reweighted trajectories using an energy bias (see Sec.~\ref{sec:infrmetad} for full details) \cite{tiwaryinfr,TiwaryE386}. 
We choose to evaluate this method first because it very quickly produces unbinding trajectories, has a metric for determining whether computed unbinding times are reliable \cite{salvalaglio}, and because we can compute free energy surfaces using standard metadynamics (MetaD) to compare the computed changes in low dimensional free energy surface with applied force to the predicted change in unbinding rate.

A constant pulling force $F$ on coordinate $Q(\vec{X})$ changes the energy of our system to $U(\vec{X})-FQ(\vec{X})$, where $Q$ is a collective variable (CV) that is a simple function of our molecular configuration $X$, such as the distance between two atoms on which we are pulling, and $U(X)$ is the potential energy of the system without an applied pulling force \cite{gomez2021molecular}.
This has the effect of ``tilting'' the probability distribution of observed configurations such that the probability of seeing some configuration at force $F$ is given by $P_F(\vec{X})=P_0(\vec{X})e^{\beta F Q(\vec{X})}$ \cite{gomez2021molecular}, where $\beta=1/(k_B T)$, $k_B$ is Boltzmann's constant, and $T$ is the temperature; $k_B T\approx{\rm 4.1 pN\ nm}$ at room temperature \cite{gomez2021molecular}. 
Because this is a static change to our probability distribution, standard equilibrium simulation techniques can be applied. 

\begin{figure}
    \includegraphics[width=\columnwidth]{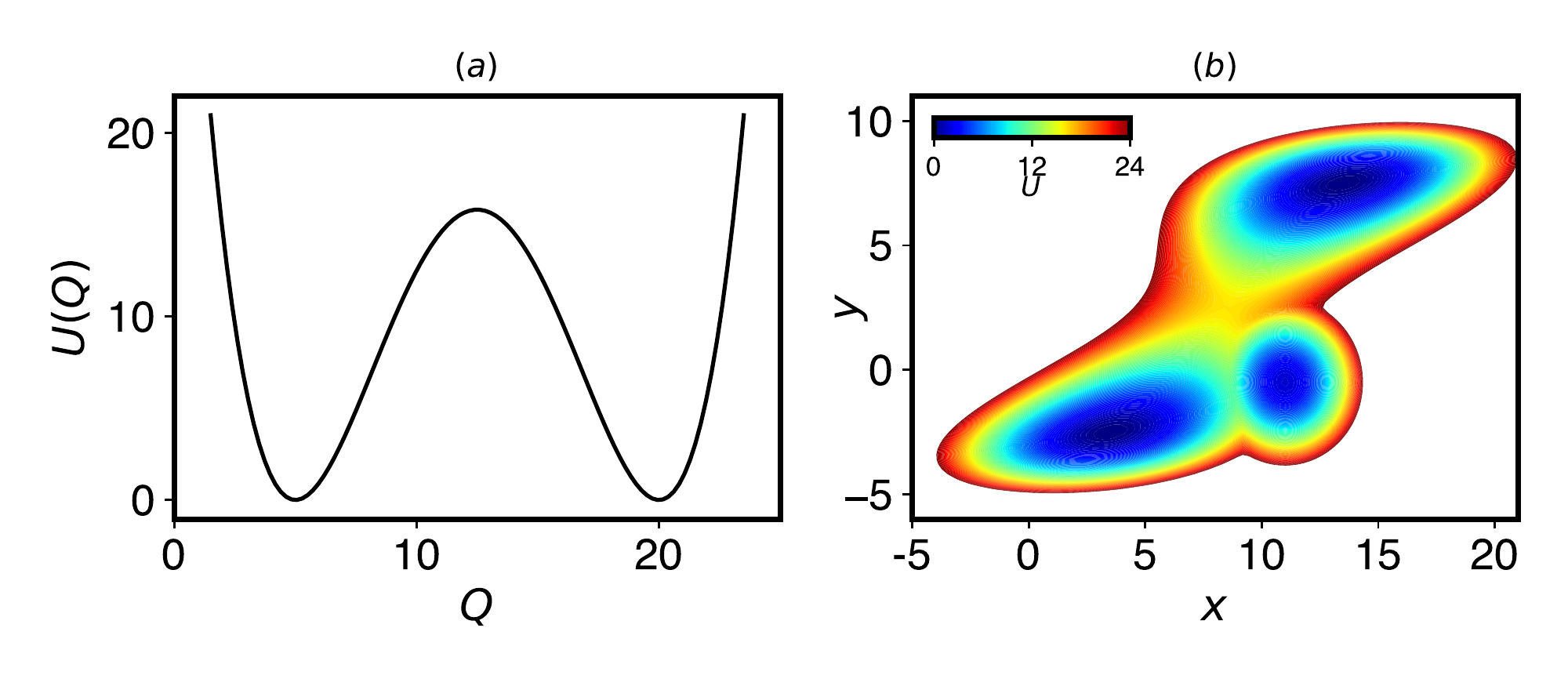}
    \caption{(a) A two well potential energy surface serves as a model system for a simple binding/unbinding reaction. (b) A three well potential energy surface, constructed by adding a metastable state to a two well potential from Ref.~\onlinecite{makarov2016perspective} is designed to exhibit catch bond behavior in escaping from the lower left to the upper right when pulling in the $x$ direction. Details of the potentials are in Sec.~\ref{sec:details_model}.}
    \label{fig:model}
\end{figure}

For a simple one dimensional energy surface such as that shown in Fig.~\ref{fig:model}a, the rate of transition from bound to unbound follows the Arrhenius law, and depends on the exponential of the height of the energy barrier between the two states.
Under certain assumptions, this implies that the dependence of an unbinding rate on force should follow 
\begin{align}
    k_{\rm off}(F)&=k_{\rm off}(0)e^{\beta F\Delta Q^{\ddag}} \label{eq:bells},
\end{align}
where $\Delta Q^\ddag$ is the distance from the bound to transition state in coordinate $Q$.
This equation was used in a theory of cellular adhesion by Bell, and hence is referred to as Bell's law in the biophysics literature \cite{Bell618}.
Bell's law is an example of a \textit{slip bond} dependence, where unbinding becomes faster with applied pulling force as we might expect. 

It is immediately obvious that the assumptions going into Bell's law need not hold for real protein systems, and hence we need not expect Bell's law to apply.
Because of this, several extended theories have been developed to correct the simplest assumptions going in to Bell's law \cite{dudko2008theory,konda2011chemical,makarov2016perspective}. 
From a broader perspective, the reason Bell's law would not hold is that the unbinding rate should depend not on the energy surface, but on the free energy surface, which at constant volume and temperature would be given by $A(Q) = -k_B T \log(\int d\vec{X} \delta(Q(\vec{X})-Q) P_F(\vec{X}))$.
Because many different molecular configurations can contribute to distances in $Q$ intermediate between bound and unbound, the free energy surface could change in unpredictable ways as force varies, and the surface may no longer be represented as a simple double well \cite{gomez2021molecular}. 

One particularly interesting class of protein-ligand bonds that we wish to study are so-called \textit{catch bonds}, where the lifetime of the protein-ligand interaction actually increases with pulling force \cite{chakrabarti2017phenomenological,thomas2008catch,barsegov2005dynamics,thomasSokurenko}. 
Physiologically, catch bonds may play many important roles, including giving cells a tool by which they can adhere strongly in the presence of strong external forces. One example is the FimH-mannose bond, which allows bacteria to adhere to the urinary tract in the presence of large shear forces \cite{thomasSokurenko,persat2015mechanical,sauer2016catch}.  
A number of theories have been put forth to explain catch bond behavior \cite{PEREVERZEV20112026,thomasSokurenko,barsegov2005dynamics,oleg}, including one type of catch bond where an applied force in one direction favors a state that has a higher barrier to unbinding (Fig.~\ref{fig:model}b).
Catch bond kinetics have not been observed directly in atomistic molecular simulation for equilibrium applied forces to date. 

An overarching question which we wish to answer in our research is, how complex does a molecular system need to be to have behavior that cannot be described by Bell's law?
Our goal here is to check whether InfrMetaD is a sufficiently powerful method to capture expected force dependent behavior for model systems where we know what the expected result should be, and then apply it to more complex molecular systems to gain insight into the molecular dissociation mechanisms that do and do not result in Bell's law behavior. 

An outline of the paper is as follows: in Sec.~\ref{sec:methods}, we describe the computational methods to be employed, including MetaD and InfrMetaD; in Sec.~\ref{sec:modelpot}, we apply these two methods to model potentials in Fig.~\ref{fig:model} and confirm that InfrMetaD can capture Bell's law and catch bond behavior; in Sec.~\ref{sec:cavity}, we then apply it to a model of a protein-receptor system, a hydrophobic ball in a hydrophobic cavity surrounded by water, and show that this exhibits Bell's law behavior, despite having a non-trivial unbinding pathway. In both cases, we evaluate the free energy surfaces to check whether their changes with force are consistent with observed differences in rates; finally, in Sec.~\ref{sec:bsa} we apply these methods to an atomistic protein-ligand system, that of biotin-streptavitin. While InfrMetaD begins to break down in this case, a combination of InfrMetaD and well-tempered MetaD suggest a number of unbinding intermediates that give rise to a breakdown in simple Bell's behavior, despite being a slip bond overall.
We discuss the ramifications of these results and the outlook for future studies in Sec.~\ref{sec:conclusions}. 
Finally, we give full details of the simulations performed above in Sec.~\ref{sec:details}.

\section{\label{sec:methods} Methods}
\subsection{\label{sec:metad}Metadynamics}
Metadynamics (MetaD) is an enhanced sampling method which allows the construction of a low dimensional free energy surface (FES) as a function of carefully chosen collective variables (CVs) \cite{metad1,bussi2020using}. An external history dependent bias that is a function of the CVs is added to the Hamiltonian of the system, pushing the system away from areas already explored \cite{bussi2020using}. 
As a result, a much wider exploration of configuration space is achieved in the same amount of MD steps. The external potential consists of a sum of Gaussians that are deposited along the trajectory of the CVs.
\begin{align}
       V(\vec{Q},t) = \omega \sum_{t'=\tau_G, 2\tau_G ...}^{t}e^{-\sum_{i=1}^d\frac{(Q_i(\vec{X}(t)) - Q_i(\vec{X}(t')))^2}{2\sigma_i^2}}
\end{align}
where $\omega$ is the Gaussian height, $\tau_G$ is the time interval at which Gaussians are deposited, $S_i$ are functions that map the atomic coordinates $\vec{X}(t)$ onto CV $i$, and $\sigma_i$ are chosen Gaussian widths for each CV.

Well-Tempered Metadynamics (WTMetaD) \cite{Barducciwt,bussi2020using} modifies the Gaussian hill heights so that they decrease exponentially as a function of the cumulative bias applied at the current CV position,
\begin{align}
    \omega'(t) = \omega e^{-\frac{V(\vec{Q},t)}{k_B\Delta T}}
\end{align}
where $\Delta T$ is the tempering factor.
Thus, as $\Delta T \rightarrow 0$, ordinary MD is recovered and as $\Delta T \rightarrow \infty$, standard MetaD is recovered.
Effectively, the CV space is sampled at temperature $T+\Delta T$, and as such WTMetaD balances an increase in the probability of crossing energy barriers with a limitation on the extent of FES exploration.
In WTMetaD, the applied bias has been shown to converge asymptotically to $-\frac{\Delta T}{T+\Delta T}F(\vec{Q})$, where $F(\vec{Q})$ is the potential of mean force \cite{dama}.

MetaD and many subsequent variations became popular for computing FESs due to its ease of use, and the fact that they promotes exploration.
As with any CV-based enhanced sampling method, the primary difficulty is choosing appropriate CVs that encompass all relevant slow transitions for the system of interest \cite{bussi2020using}.

\subsection{\label{sec:infrmetad}Infrequent metadynamics}
Although MetaD was designed to predict static properties of a system such as the FES, in some situations it can be adapted to produce an estimate of the rate of slow dynamical events \cite{tiwaryinfr}.
Voter demonstrated that unbiased rates of infrequent barrier crossing processes can be computed very rapidly by applying a bias outside of transition regions to ``boost'' the system over those barriers \cite{voter,voter2}.
Tiwary and Parrinello proposed the idea of infrequent metadynamics (InfrMetaD), where the metadynamics framework described above is used to produce this boost potential on the fly \cite{tiwaryinfr}. 
In order to extract unbiased rates, three key conditions must be met: (1) the transitions from one state to the other are rare, but the actual crossing of the transition state is ephemeral, (2) the biased CV is a good reaction coordinate for the transition, and (3) additional Gaussians are added to the bias potential infrequently enough that none are added during the barrier crossing \cite{tiwaryinfr}.

When this is the case, transition state theory says that the ratio of the escape times in the biased and unbiased cases is given by the ratio of the partition functions in the reactant basin. This ratio gives an acceleration factor $\alpha$ which can be computed as 
\begin{equation}
    {\displaystyle \alpha} = {\langle \displaystyle e^{\beta V(\vec{Q},t)} \rangle} \label{meq7}
\end{equation} 
where $\vec{Q}$ are the collective variables being infrequently biased and $V(\vec{Q},t)$ is the metadynamics bias experienced at time $t$ \cite{tiwaryinfr}.  

In order to estimate a rate using InfrMetaD, many trajectories are run with different random seeds up to the points where the system was deemed to have reached the unbound state.
The unbiased reaction times for each simulation instance are estimated by multiplying the final time in the simulation by the acceleration factor computed up to that point. 
Because unbinding is a rare event, we expect the distribution of transition times to be exponential as for a homogeneous Poisson processes, and to depend on a single bond lifetime $\tau$ \cite{Tiwary12015,salvalaglio}. 
Hence, to obtain the unbinding rate, a cumulative distribution function (CDF) from computed unbiased transition times can be built and fit to the ideal CDF,
\begin{align}
   CDF(t) =  1 - e^{-t/\tau}. \label{eq11}
\end{align}
The rate of the process can then be computed as $k=1/\tau$.
The correspondence of the empirical CDF to the ideal CDF can be checked by the Kolmogorov-Smirnov (K-S) test \cite{salvalaglio}, which we do with the python package scipy.stats \cite{virtanen2020scipy}.

\section{Results and discussion}
\label{sec:results}

\subsection{Model potentials}
\label{sec:modelpot}
We first wish to confirm that InfrMetaD is an appropriate tool to predict the force dependence of unbinding rates. 
To do this, we apply InfrMetaD to one- and two-dimensional potentials meant to exhibit slip and catch bond behavior, respectively.
As described in Sec.~\ref{sec:intro}, a two well potential such as that in  Fig.~\ref{fig:model}a is predicted to show Bell's law dependence of unbinding rate with force. 

To compute the unbinding rates, a total of 20 InfrMetaD runs were performed for each force, where pulling and bias were applied to the $x$ coordinate (see Sec.~\ref{sec:details_model}, with representative CDFs in Fig.~\ref{fig:cdfslip}).
Moreover, we explicitly compute the free energy using WTMetaD to see how the change in the underlying FES corresponds to the change in rate.
Here, we do this to be consistent with forthcoming examples and to verify our numerical approaches, although it is not necessary for a one-dimensional case.

\begin{figure}[ht]
\includegraphics[width=\columnwidth]{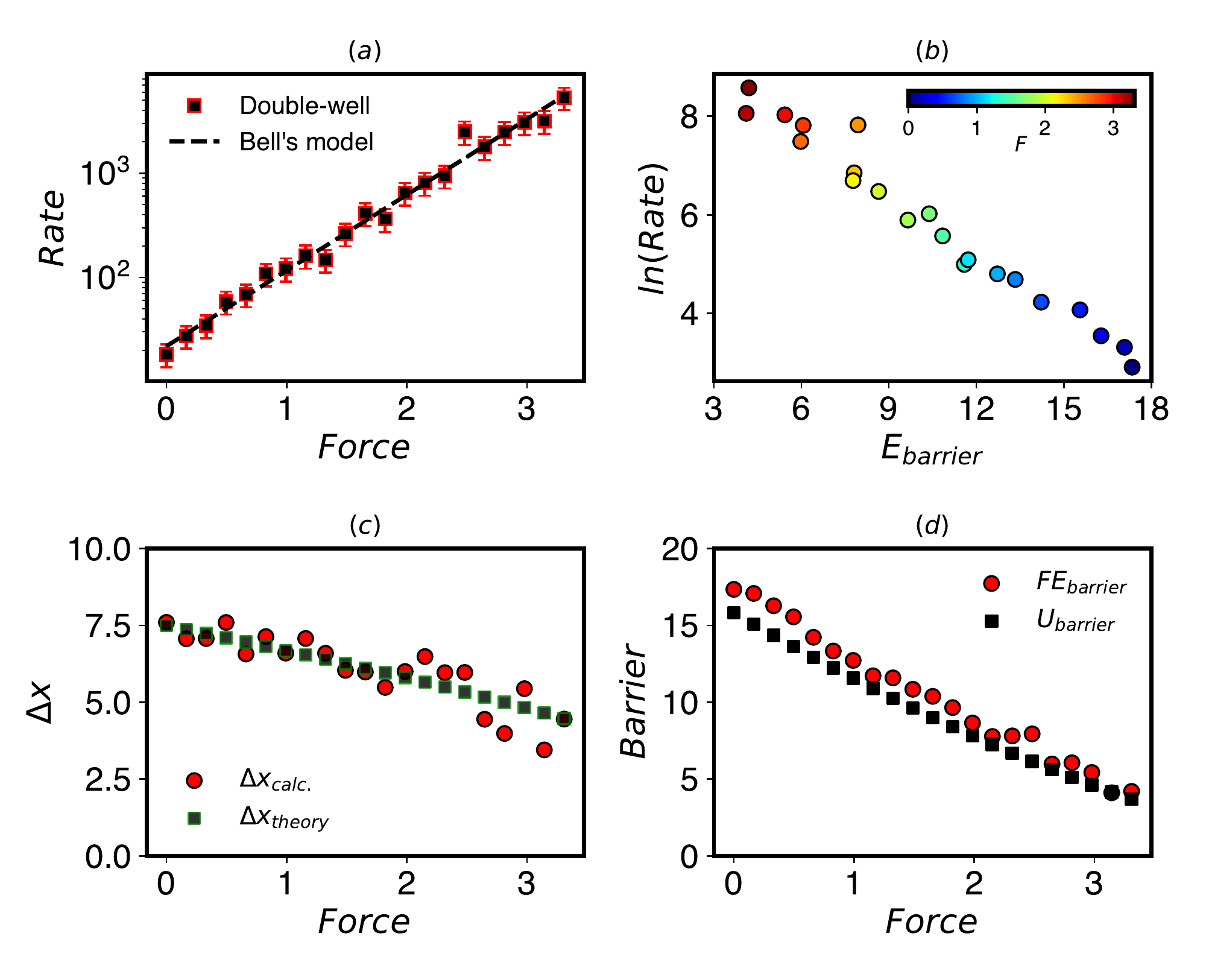}
\caption{\label{fig:pot1d} (a) Unbinding rates computed by InfrMetaD for the potential in Fig.~\ref{fig:model}a. The log of the rates increases linearly with force, apparently following Bell's law. The fit parameters are $k^0$ = 21.90, and $\Delta x^{\ddag}$=6.86. (b) Rates computed by InfrMetaD plotted against the energy barrier computed by WTMetaD exhibit Arrhenius behavior. (c) The transition distances computed from FE calculations shrinks with applied force as predicted by extended Bell's theory \cite{konda2011chemical}, meaning that not all assumptions of Bell's law are true. ``Theory'' values are the shift in the analytical potential with force. (d) The energy barrier to unbinding decreases linearly as higher forces are applied, in accordance with the assumption going in to Bell's law. $U_{\rm barrier}$ is the analytical barrier height.}
\end{figure}

As expected, the rates computed by this approach increase exponentially with applied force, and fit very well to Bell's law (Fig.~\ref{fig:pot1d}a). 
How does this connect to the underlying (free) energy surface?
Fig.~\ref{fig:pot1d}b shows that these rates conform to the Arrhenius law, where the rates are exponentially dependent on the barrier height between the states.
Among other assumptions, Bell's law should hold when the barrier decreases linearly with force and the distance to the transition state is constant \cite{gomez2021molecular}. 
Fig.~\ref{fig:pot1d}c-d shows that the agreement with Bell's law is a bit fortuitous, because it exhibits (an expected) linear shift of the transition state distance with applied force that is taken into account using an extension of Bell's law \cite{konda2011chemical}.

We now move beyond this trivial first test to assess whether InfrMetaD can capture catch bond behavior in a model system.
The catch bond potential we have created is adapted from Ref.~\onlinecite{makarov2016perspective}, but we have added a third potential well that has a higher transition barrier to the product (top right) state (Fig.~\ref{fig:model}b). 
We predict that upon pulling to the right in $x$, the intermediate will be stabilized, and the barriers will change such that the most favorable path is through the intermediate, which still has a slower rate of transition to the product.

\begin{figure}[ht]
\includegraphics[width=2.2in]{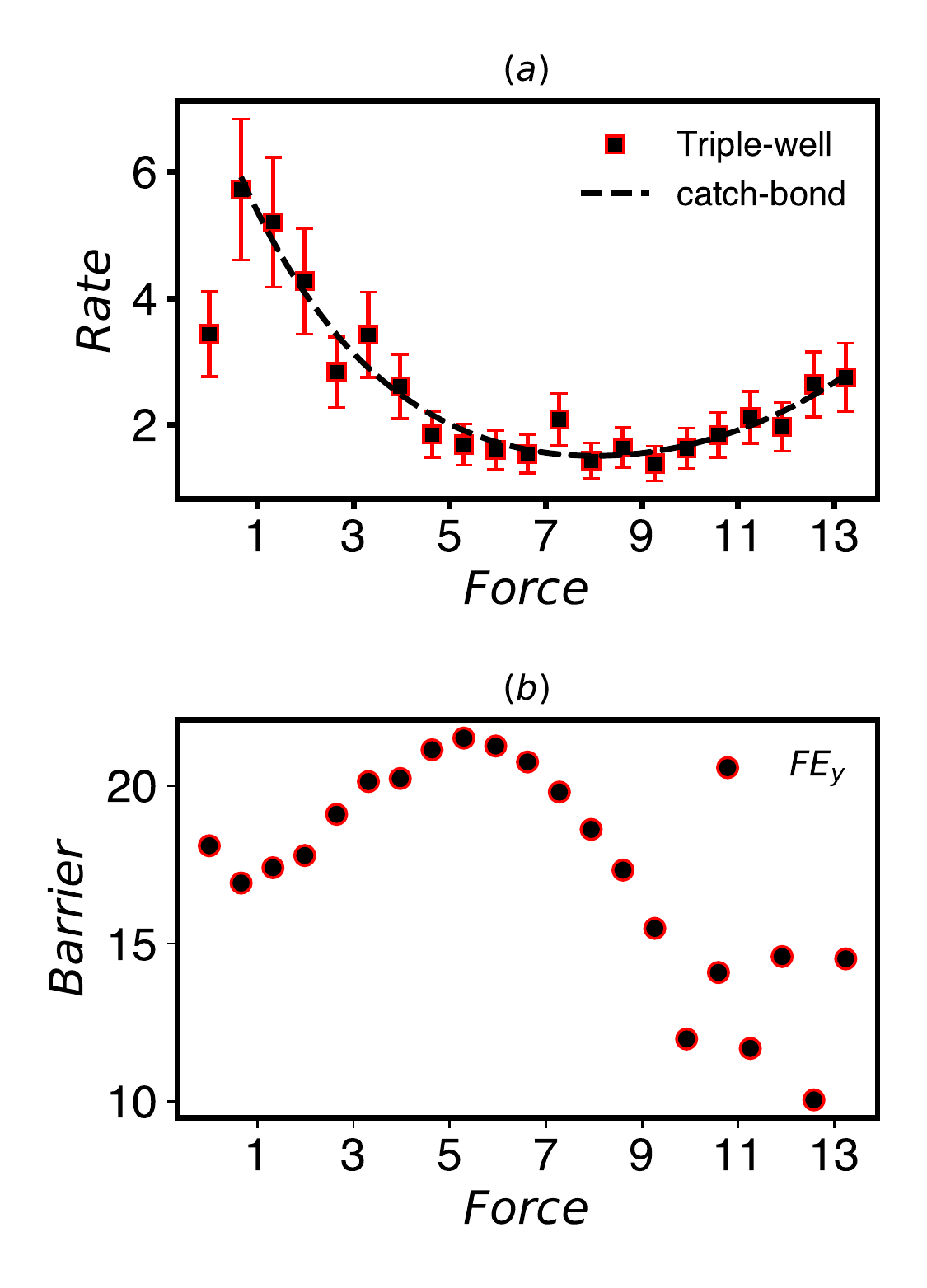}
\caption{\label{fig:pot2drate} (a) Rates computed by InfrMetaD for the potential 
in Fig.~\ref{fig:model}b. This system exhibits catch-slip bond behavior, and the rate dependence can be fit well to a catch-slip rate dependence (dashed line) as described in the main text. (b) The free energy barrier in the $y$ direction computed from data in Fig.~\ref{fig:fes2d} using Eq.~\ref{eq:intout} shows an increase and decrease with force mostly commensurate with the rate dependence. However, the barrier apparently begins to decrease before the rate begins increasing, showing that explaining the full dynamics requires knowledge of the full 2d surface.}
\end{figure}

To compute the rates, 20 InfrMetaD runs were performed for each force applied in the $x$ direction, while the WTMetaD bias is applied symmetrically in both the $x$ and $y$ coordinates (see Sec.~\ref{sec:details_model} for simulation details, with representative CDFs in Fig.~\ref{fig:cdfcatch}). 
We observe for this model that the rate of unbinding decreases in the range $F\in\{1,8\}$ and then increases from that point onward, an example of a catch-slip bond (Fig.~\ref{fig:pot2drate}a). 
The existence of the intermediate state causes the rate dependence to deviate from Bell's model except at the very smallest forces.
We can fit the observed behavior well using a sum-of-exponential catch-slip rate dependence \cite{oleg} given by $k_{\rm bottom\rightarrow top} = k_ce^{-x_cF\beta} + k_se^{x_sF\beta}$ where $k_c$, $x_c$, $k_s$ and $x_s$ have values of 7.00, 1.24, 0.17, and 0.86 respectively for the curve in Fig.~\ref{fig:pot2drate}a.

\begin{figure}[ht]
\includegraphics[width=\columnwidth]{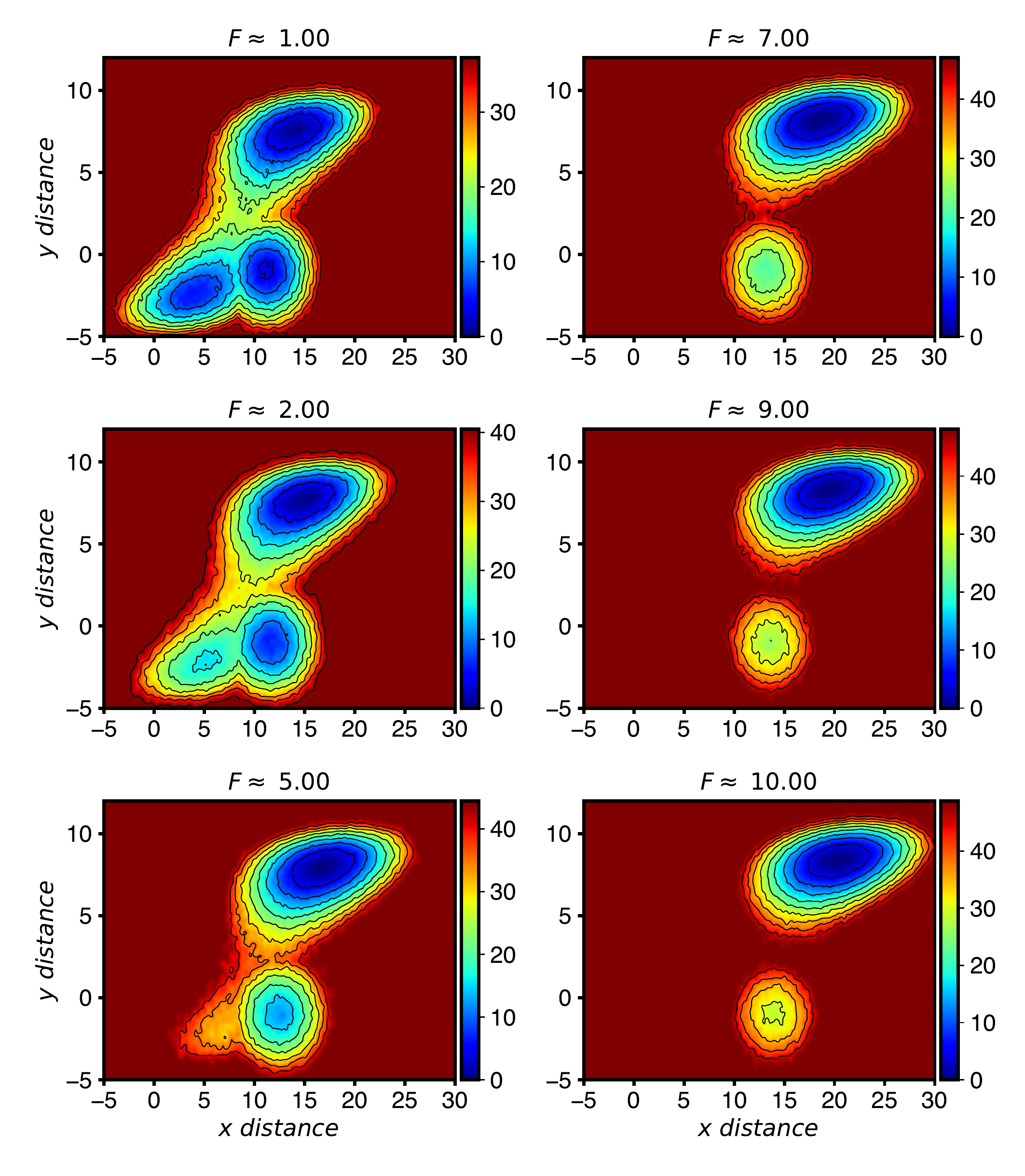}
\caption{\label{fig:fes2d} Free energy surface for the potential in Fig.~\ref{fig:model}b, at different pulling forces. Pulling in the $x$ direction lowers the system's energy proportional to its $x$ location. This causes the upper state which is farthest to the right to become the dominant state at higher forces, and also causes the lower-left state to become unstable and vanish.}
\end{figure}

We next use WTMetaD to check our intuition for how the free energy surface is changing. 
Our results in Fig.~\ref{fig:fes2d} show that the situation is similar but more complex than our initial expectations. 
At small force, it can be seen that all or most MetaD transitions took place directly between the lower and upper state. 
At $F=2$ and $F=4.6$, the force in the $x$ direction makes the intermediate state lower in free energy, which has a higher barrier to escape. 
Between $F=5$ and $F=7$, the original stable state has vanished.
The rate is still decreasing and the barrier increasing, but this is due to the shift in relative positions of the two minima. 
It is only once the upper state is fully to the right of the initial intermediate just above $F=7$ that the unbinding rate starts to increase again.

An effective one dimensional free energy surface in the $y$ direction, $A(y)$ can be computed by integrating out the $x$ dependence,
\begin{equation}
    A(y) = \int_{-\infty}^{\infty} e^{-\beta A(x,y)}dx
    \label{eq:intout}
\end{equation}
The transition barrier between states for $A(y)$ is shown in Fig.~\ref{fig:pot2drate}b.
Here, the change in barrier in the $y$ direction is mostly consistent with the change in observed rates, however there is some disagreement in the range $F=5-7$, where the barrier goes down but the rate continues decreasing with increased force. 
This is a prime example of how projecting to a low dimensional free energy surface can hide important structural information that affects the prediction of rates \cite{makarov2016perspective}, motivating the use of methods to directly compute rates rather than trying to infer rates from computed FESs.

\subsection{Cavity ligand model}
\label{sec:cavity}
Having demonstrated that InfrMetaD and also WTMetaD are capable of extracting the force dependence of unbinding in accordance with our expectations, we now turn to an explicit, all atom but simplified representation of a ligand unbinding process---a hydrophobic sphere contained in a hydrophobic cavity, solvated by water \cite{mondal2013hydrophobic,Tiwary12015,mondal2020}.

\begin{figure}[ht]
\includegraphics[width=5cm]{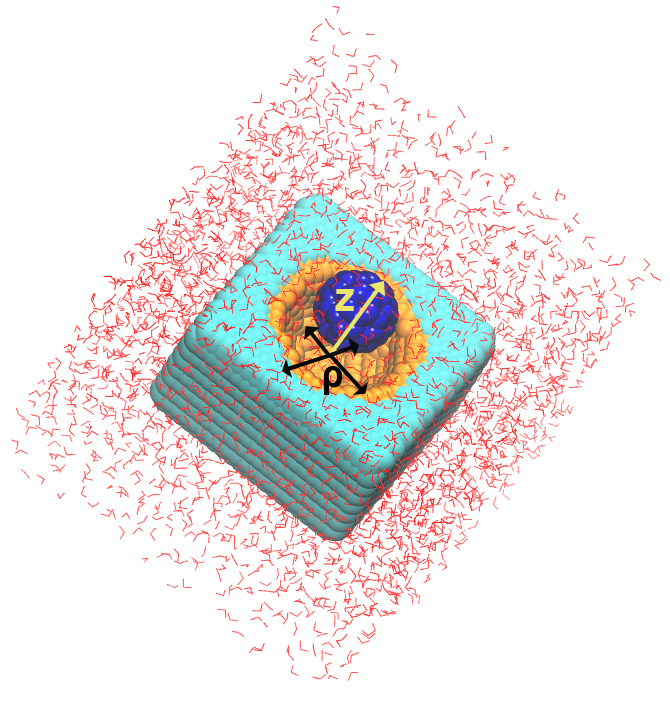}
\caption{\label{fig:cavityimg} Solvated cavity-ligand model from Ref.~\onlinecite{mondal2020}. The cyan and orange atoms (CW, CP) form the receptor and the blue atoms (CF) make up the ligand. The radial and perpendicular distance CVs used for biasing are labeled on the figure.}
\end{figure}

We choose this model for two reasons: (1) through extensive studies, it is known that the unbinding pathway for this system involves first moving sideways before exiting, because a direct perpendicular exist requires water molecules to fill in a vacuum created by the fluctuation of the sphere out of the cavity.
This means that the unbinding process is not well described by considering the obvious obvious reaction and pulling coordinate (central distance of the ball from the cavity) \cite{mondal2013hydrophobic,Tiwary12015,mondal2020}, and (2) this system has been well characterized at zero force by both WTMetaD and InfrMetaD, hence we expect our calculations to be converged using the same protocols. 

\begin{figure}[ht]
  \includegraphics[width=\columnwidth]{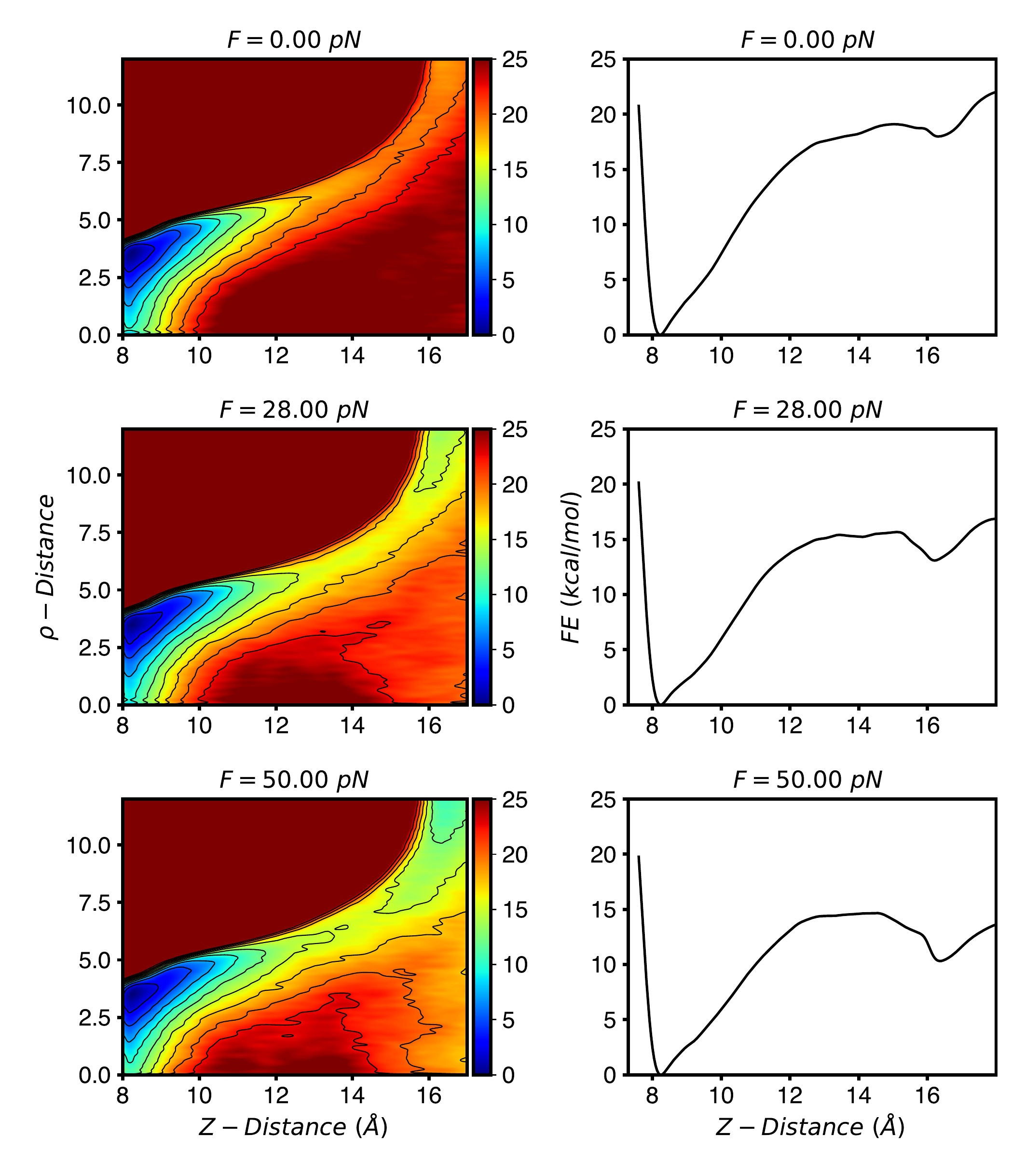}
  \caption{(Left) FESs computed for the cavity-ligand system (Fig.~\ref{fig:cavityimg}) at three different forces. Pulling along a central distance coordinate increases stability of the unbound state. (Right) One dimensional free energy surfaces computed by integrating out the $\rho$-distance according to Eq.~\ref{eq:intout}.}
  \label{fig:cavityfes}
\end{figure}

In order to perform InfrMetaD computations for this system, we should have a good estimate of a distance that we consider the sphere to be unbound. 
Here, we first performed WTMetaD calculations on this model with different applied forces, using the same protocol as Ref.~\onlinecite{mondal2020} (full details in Sec.~\ref{sec:details_cavity}).
FESs at $T=300$K were obtained for forces $F$=0 to 50 pN, in 2 pN intervals, where pulling forces are applied to the full three-dimensional distance between the center of mass of the cavity and the center of mass of the sphere. 
The MetaD bias was applied to two CVs, the radial and transverse distance of the sphere from the center of the cavity (Fig.~\ref{fig:cavityimg}).

Fig.~\ref{fig:cavityfes}(left) shows the computed FES in our two CVs at three different forces.
As described in previous work, the FES at zero force clearly shows that the escape of the sphere involves a radial shift away from the central axis before exiting, which more easily allows water into the cavity \cite{mondal2013hydrophobic,Tiwary12015,mondal2020}. 
When projected in just the transverse direction ($Z$), Fig.~\ref{fig:cavityfes}(right), we see that the FESs resembles a prototypical double well potential. 
Application of a pulling force lowers the free energy of the unbound state, as well as the barrier between the bound and unbound state. 

\begin{figure}[ht]
\includegraphics[width=\columnwidth]{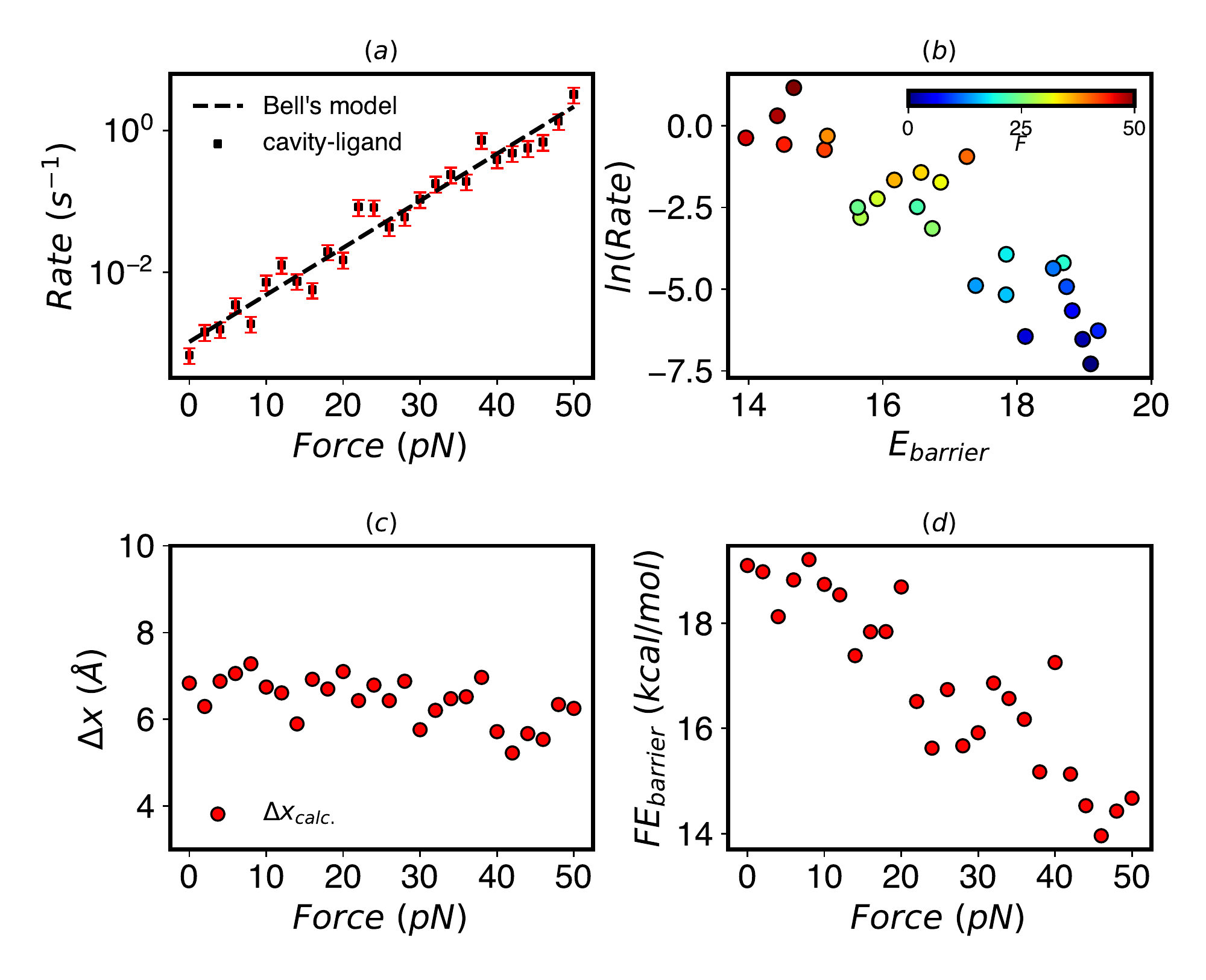}
\caption{\label{fig:cavityrate} Cavity-ligand model. (a) Unbinding rates from InfrMetaD increase exponentially with pulling forces. Fit parameters to Bell's law are $k^0$=0.0011 $s^{-1}$ and $\Delta x^{\ddag}$=6.31 \AA . (b) The rate decreases exponentially with barrier height computed from WTMetaD projected along the $Z$-distance using Eq.~\ref{eq:intout}. (c) The distance to the transition state in the $Z$ direction decreases slightly but is relatively constant compared to the two well potential results in Fig.~\ref{fig:pot1d}c. (d) The free energy barrier in the $Z$ direction decreases linearly with force. }
\end{figure}

While these FESs can help us understand the mechanism of unbinding at different forces, they do not give us direct access to the unbinding rates. 
Now that we know the transition distance, approximately 15 \AA\ in all cases, we can apply InfrMetaD to this system (see Sec.~\ref{sec:details_cavity} for full details, with representative CDFs in Fig.~\ref{fig:cdfcavity}).
Although the unbinding process is much more complicated than for a 2-well potential, we observe in Fig.~\ref{fig:cavityrate}(a) that unbinding rates increase exponentially with increasing force.

Again combining our data from WTMetaD and InfrMetaD, we show that unbinding rates for this cavity model fit well to the Arrhenius law across our range of forces, but not nearly so well as for a true one dimensional double well (Figure \ref{fig:cavityrate}d).
Interestingly, the transition distance in the 1-dimensionalized potential is almost constant, as can be seen in Fig.~\ref{fig:cavityrate}c, while the barrier shown in (d) decreases linearly. 
Despite its non-trivial unbinding pathway, this cavity-ligand model is closer to ideal Bell's law behavior than the 1d potential upon which the theory is based.
We speculate that this is due to the rigidity of the cavity and ligand, and this relationship could begin to break down if the cavity were made more flexible.

\subsection{Fully atomistic protein-ligand system}
\label{sec:bsa}
\begin{figure}[ht]
\includegraphics[width=8cm]{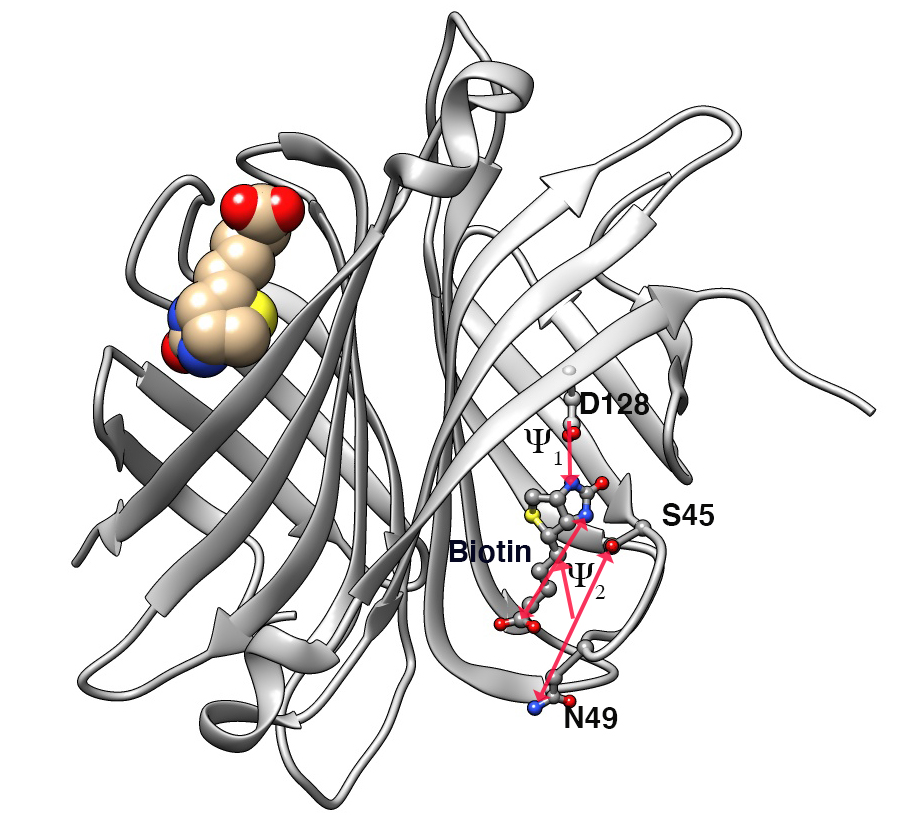}
\caption{\label{fig:fig:bsaimg} Dimeric form of Streptavidin in complex with biotin. Only one biotin was chosen to perform InfrMetaD. The other biotin molecule remained in its bound pose throughout the duration of all simulations. Protein residues and distances going into the bias CV are labeled.}
\end{figure}

Given the reasonableness of our prior results on the cavity-ligand model, we sought to apply our approach to a fully atomistic protein-ligand system.
We chose to study the biotin-streptavidin (SA/b) bond (Fig.~\ref{fig:fig:bsaimg}) for three reasons: (1) it plays an important role in many \textit{in vitro } biochemical studies and is one of the strongest biological non-covalent bonds known, \cite{Rico6594} (2) its bond rupture has been studied in non-equilibrium pulling experiments and simulations \cite{Rico6594}, and (3) its unbinding kinetics at zero force have been assessed previously using InfrMetaD \cite{tiwarybio}.

Computing unbinding rates of protein-ligand systems is an active area of research and is clearly non-trivial. 
A major challenge, as discussed above, is choosing a good reaction coordinate. In Ref.~\onlinecite{tiwarybio}, Tiwary optimized a slow reaction coordinate using the SGOOP algorithm \cite{tiwary2016spectral} for the unbinding of the biotin ligand, which is a linear combination of distances between the ligand and residues in the binding pocket (Fig.~\ref{fig:fig:bsaimg}).
This optimized coordinate allowed InfrMetaD unbinding times to pass the statistical test, although there are signatures of non-exponential behavior in the data attributed to metastable intermediates along the unbinding pathway (seen also in Ref.~\onlinecite{Rico6594}). 

\begin{figure}[ht]
\includegraphics[width=\columnwidth]{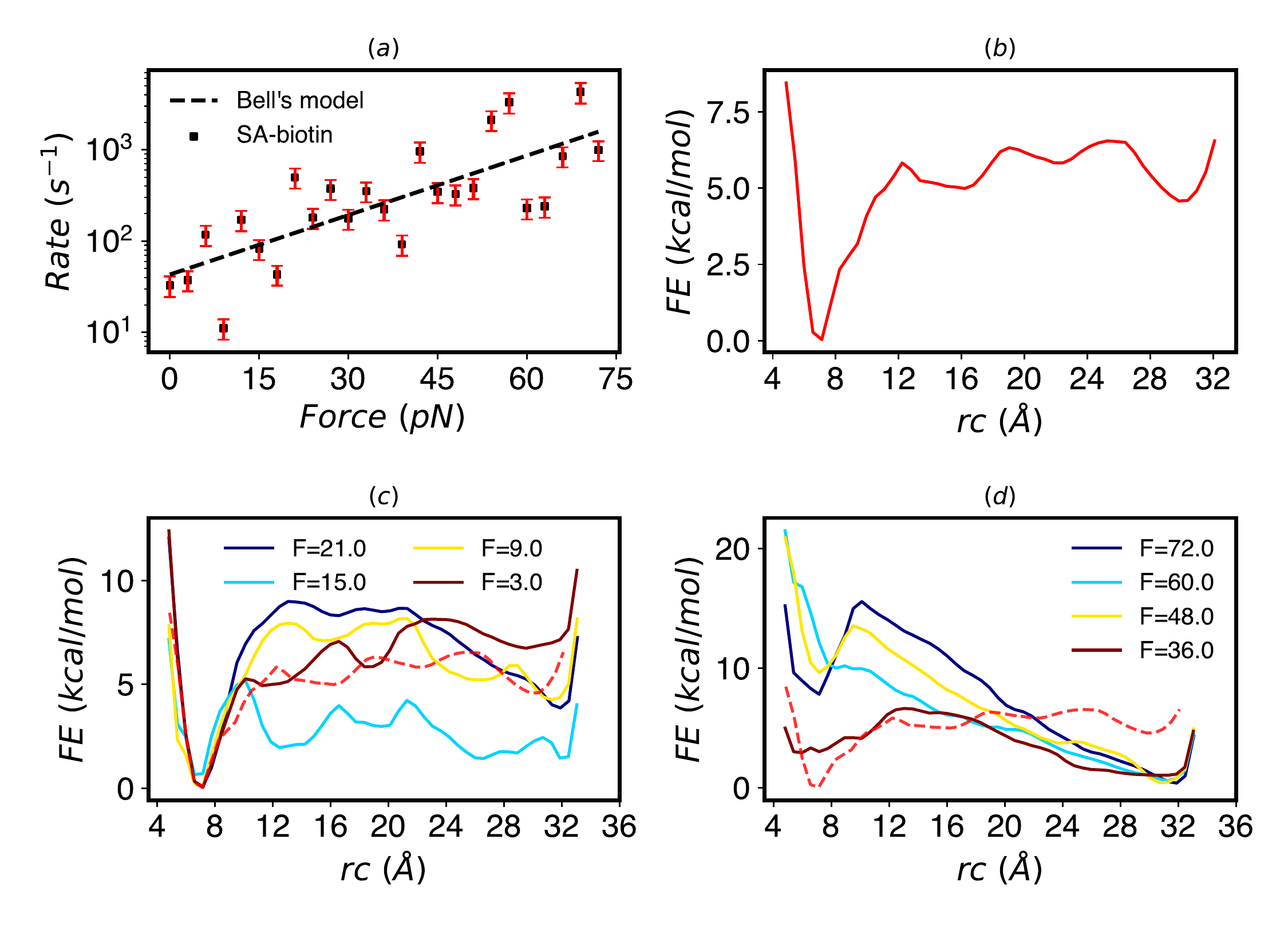}
\caption{\label{fig:bsalograte} (a) Unbinding rates computed from InfrMetaD show an overall slip bond, but a poor fit to Bell's law. The fit parameters used are $\Delta rc^{\ddag}$=2.1 \AA, $k^0$=43 $s^{-1}$. This fitted $\Delta rc^{\ddag}$ is not close to the transition distance observed in FES computations. (b) An approximate FES for the system at zero force shows two unbinding intermediates along the optimized MetaD coordinate. (c) Smaller forces shift the energy of the intermediate states but the shifts seen are not monotonic as $F$ increases. (d) When applying larger forces, the computed FES shows a loss of metastable intermediates and a shift to a favored unbound state.}
\end{figure}

Following Ref.~\onlinecite{tiwarybio}, we constructed a dimeric SA/b complex, and studied the unbinding of one of the two biotin ligands using InfrMetaD. With slightly different MetaD parameters (bias was deposited more infrequently, every 15 ps rather than 5 ps), we get an unbinding rate of $32.66 \pm 8.22 s^{-1}$ which as in Ref.~\onlinecite{tiwarybio} is much faster than the measured rate for the full tetrameric complex. 
We then proceeded to compute the unbinding rates as a function of force, with 20 InfrMetaD runs performed for forces in the range $F=0$ pN to 72 pN.
Unfortunately, despite numerous attempts to adjust the InfrMetaD pace, hill height, and width, we were unable to obtain unbinding rate distributions that pass the statistical tests for most forces.  
The rates obtained from the parameters that gave our closest to exponential results are shown in Fig.~\ref{fig:bsalograte}, with representative CDFs that do an do not pass the KS test shown in Fig.~\ref{fig:cdfbsa}.

Despite the fact that we cannot say with confidence that these unbinding times are accurate or converged, they are quite consistent with experimental results measured by dynamic force spectroscopy in Ref.~\onlinecite{Rico6594}, which vary from approximately 30-1000 $s^{-1}$ as force ranges from 0 to 75 pN (see Ref.~\onlinecite{Rico6594} Fig. 3E). 
Overall, our predicted rates follow a Bell's law like trend, however there are substantial deviations from the trend, which coincide with what is likely a much more complex unbinding energy landscape in this case.
It is unclear whether substantial dips at 9 pN and 18 pN could correspond to any catch-bond like behavior, or are simply an indication that our computations are not well converged. 

In order to gain some insight into the reason the InfrMetaD breaks down, we compute an approximate FE surface using WTMetaD, while restricting the ligand to stay close to its initial monomer using a `wall' constraint (see Sec.~\ref{sec:details_bsa} for full details). These FE surfaces in Fig.~\ref{fig:bsalograte}b-d reveal multiple unbinding intermediates, as suggested in Ref.~\onlinecite{tiwarybio,Rico6594}. 
Here, we can clearly see that the roughness of the surface becomes more pronounced for intermediate forces. 
At larger forces in Fig.~\ref{fig:bsalograte}d, the surface becomes more smooth again and the unbound state is clearly favored.
Similar free energy surfaces projected on the pulling coordinate, which is what we would normally tend to show to get insight into the unbinding process, are shown in Fig.~\ref{fig:fesbsapull}. 
However, showing the surface in terms of the reaction coordinate used in InfrMetaD is more appropriate for diagnosing why the assumptions going in to the rate computations are not being satisfied.

In total, these MetaD and InfrMetaD data together suggest that while the coordinate obtained from Ref.~\onlinecite{tiwarybio} was apparently good enough for use at $F=0$ pN, it is not sufficiently optimized for higher forces. 
This poses a challenge going forward as to whether a single CV or set of CVs can be determined that is appropriate for all forces, or whether a new reaction coordinate must be determined for each pulling force, since the application of force can change the underlying free energy landscape in unpredictable ways.

\section{\label{sec:conclusions}Conclusion}
Thermodynamic and kinetics calculations were performed for various models with increasing complexity to determine the force dependence of transition rates.
In the case of a simple two-well potential and a hydrophobic ball/cavity system, we showed that unbinding rates increased exponentially with force, while a model catch bond system showed a decrease in unbinding rate corresponding to stabilization of an intermediate. 
For the biotin-streptavidin interaction, the presence of multiple intermediates causes InfrMetaD to break down, as the unbinding process is no longer a single high energy barrier (using the chosen bias CV). 
Intriguingly, our rough results for the free energy surface from MetaD show very non-monotonic changes with force that could be indications of metastable unbinding states stabilized by the applied force.
Our results also suggest, as described in earlier work, that the failure of unbinding time cumulative distributions to be exponential are reflective of the complexity of the unbinding pathway, and can be used to help diagnose whether a good reaction coordinate has been chosen for InfrMetaD in the presence of force.  

We believe that our prediction that the hydrophobic cavity system exhibits true Bell's law behavior is the first such explicit prediction from equilibrium MD simulations. 
The concordance between MetaD and InfrMetaD results and the relative efficiency of InfrMetaD do suggest that InfrMetaD is a promising technique to evaluate the force-dependence of unbinding rates for complex systems. 
However, its failure to past statistical sanity checks for most forces in the case of SA/b serves as a warning to those, including us, who hope to apply such techniques to even more complex systems, such as large protein-protein complexes that exhibit catch bond behavior. 
In our case, we chose the SA/b system because an optimized coordinate had been previously computed for use in InfrMetaD. 
Yet this coordinate was insufficient once forces were applied. One possible solution to this problem is to compute new reaction coordinates for each applied force.
Although this would be cumbersome, it is certainly a more rigorous approach that we will explore in subsequent work. 

Given our current approach though, it remains to be seen whether the overall trend of increasing or decreasing rates computed from InfrMetaD, even in the presence of this break-down, could be a fingerprint of catch or slip bond behavior, and give some insight into the mechanisms. 
This would be analogous to the ways in which steered-MD has given important insights into unbinding or unfolding reaction mechanisms despite generally producing unrealistically high forces at unrealistically fast rates. 

Finally, in the the future we plan to explore other equilibrium methods for computing rates, to determine whether they are more suitable for computing force dependence. 
We are currently evaluating the weighted ensemble approach, which has the advantage of not having to choose a specific reaction coordinate on which to apply an energy bias, although a choice of a progress coordinate for unbinding is still needed. 
Because this method requires performing many cycles of simulation, we can actually predict that it will become more computationally effective for the case of slip bonds, as applied forces will result in faster rates, which requires fewer cycles to converge. 
At the same time, catch bonds will require a higher cost than computing a zero force unbinding rate, due to the longer lifetime of the bond. 
We plan to present a detailed comparison of these approaches in the near future. 

\section{\label{sec:details}Simulation Details}

\subsection{Pulling}
\label{sec:details_pulling}
The PLUMED plugin library \cite{plumed1,plumed2} was used to apply WTMetaD and pulling forces. 
A pulling force is achieved via a bias generated by a linear restraint formulated in PLUMED as: 
\begin{equation}
    U_{\rm external}=F(Q-a) \label{meq8}
\end{equation} 
where $F$ is a force constant in units of energy over length in $Q$ units, $Q$ is the CV to which the force is being applied and $a$ is the location of the restraint, which only sets the zero of energy but does not change the force applied.
Therefore, in order to apply pulling forces, a negative $F$ is fed to PLUMED.
For our atomistic simulations, $Q$ will be some distance in the protein, and units of kcal/mol and \AA\ will be used, and so forces must be applied in units of 1 kcal/(mol \AA). 
Piconewtons can be computed into this unit system with $\approx 69.48\ {\rm pN}$ equivalent to 1 kcal/(mol\ \AA).

\subsection{\label{sec:level5}Rate Calculation}
To compute rates, many simulations must be run for each pulling force. 
Simulations are run up to the point where the ligand reaches the unbound state. The COMMITTOR feature of PLUMED is used to terminate the simulation once an unbinding coordinate reaches a specified value.  
A WTMetaD bias is applied using the METAD feature of PLUMED using the ACCELERATION keyword, such that $\alpha$ is computed within each simulation.
The time at which the unbound state is reached and the acceleration factor are recorded for each run. The product of the simulation time and acceleration factor gives the scaled residence time for each run. These sets of scaled transition times were used to determine the mean residence time and unbinding rate for a given force using the following procedure. 

ECDFs were built by histogramming the transition times against a set of log-spaced bins and getting a cumulative sum of the histogram divided by the total number of transition times. To determine how well the assumptions of InfrMetaD were met during the simulations and to validate computed rates, the protocol of Ref.~\onlinecite{salvalaglio} was followed. The two sample Kolmogorov-Smirnoff test determines the similarity between the transition time distributions. The null hypothesis is rejected if the transition times obtained from MetaD and the transition times obtained randomly from an exponential CDF with corresponding $\tau$ parameter do not come from the same underlying distribution at the 5 \% significance level. Additionally, the p-value provides a measure of goodness of fit. If the p-value of the KS test is higher than 0.05 then the sample distributions are said to come from the same underlying distribution.

\subsubsection{Model potentials}
\label{sec:details_model}
Model systems from Fig.~\ref{fig:model} were simulated using the PESMD feature of PLUMED.
The potential in Fig.~\ref{fig:model}a is given by $U(x)=0.005 (x-5)^2(x-20)^2$, with a a ``bound'' state at $x=5$ an ``unbound'' state at $x=20$  separated by an energy barrier of 15.8 $k_B T$. 
The potential in Fig.~\ref{fig:model}b is constructed as a Gaussian mixture model, combining the two well potential from Ref.~\onlinecite{makarov2016perspective} with an additional harmonic potential.
The potential energy is given by 
\begin{align}
\begin{split}
    U(x,y)=&-\ln(e^{-((0.4y-1)^2)-4)^2+\frac{1}{2}(x-6-y)^2} \\
    & + 0.2e^{-(x-12)^2-2(y+0.5)^2})
\end{split}
\end{align}
This results in minima  at (3.5, -2.5), (13.5, 7.5) and (11, -0.5) representing the bound, unbound and p-bound state respectively (p-bound refers to bound by pulling). The bound and unbound state are separated by a high energy barrier of 16 $k_B T$. The barrier between the p-bound state and unbound state is higher than from the bound state, 21  $k_B T$. 
For the double well potential InfrMetaD was performed with HEIGHT=1.2, SIGMA=0.2 BIASFACTOR=6 and PACE=4500. A total of 20 simulations were run for each pulling force; simulations were run up to the point at which the unbinding CV reached the unbound state. 
FES calculations were performed with WTmetaD for every pulling force using the same parameters as above except for PACE which was set to 650 and these simulations performed for $7.5\times10^6$ MD steps.  
For the three well potential InfrMetaD was performed with HEIGHT=1.2, SIGMA=0.2,0.2 BIASFACTOR=6 and PACE=7500, both the x and y component of the distance were biased but the pulling force was applied only in the x direction. Similarly here, a total of 20 simulations were run at each pulling force until the unbound state was reached. 
FES calculations were performed with WTMetaD for every force using the same parameters as above except for PACE which was set to 600. These simulations were performed for 2$\times10^7$ MD steps. The default PLUMED units were used for all simulations of the model potentials. The simulation time step used in both cases was $dt=0.002$.

\subsubsection{Cavity-ligand model}
\label{sec:details_cavity}
The model consists of a semi-hollow cube of pseudo atoms resembling carbon atoms that are ordered in a hexagonally close-packed lattice. Moreover, the cube consists of two categories of hydrophobic atoms; the cavity atoms and the anchor or wall atoms. The radius of the cavity is 8 \AA, and the lattice constant is $a$ 2 \AA. 
The ligand is a sixty-atom (C60) fullerene (bucky ball), which has a weak van der Waals atraction to the cavity.  The atoms in the cavity have a higher attraction to the ligand than do the anchor atoms and the whole complex model is solvated with TIP4P water (Fig.~\ref{fig:cavityimg}). GROMACS \cite{gromacs} files for this model from Ref.~\onlinecite{mondal2020} were provided by the Mondal group.

Non-bonded interactions are determined by GROMACS using the OPLS combination rule. The non-bonded interactions of the CP-CP, CW-CP, and CW-CW pairs were excluded by setting their LJ parameters to 0. The entire lattice was fixed in position. The interaction between the different molecules other than water are summarized in Table \ref{tab:lj}.

InfrMetaD was performed using HEIGHT=0.287, SIGMA=0.3 BIASFACTOR=15 and PACE=5000. Both the InfrMetaD bias and the pulling force were applied on the 3D distance between the COM of the ligand and the COM of the cavity. A set of 20 simulations were run for each pulling force until the unbound state was reached.

FES calculations were performed for all forces using HEIGHT=0.478, SIGMA=0.3,0.1 BIASFACTOR=15 and PACE=300. The bias was applied to both the transverse and radial distance of the sphere from the center of the cavity respectively while the pulling force was applied to the 3D distance. PLUMED walls were applied for the transverse and radial distance CVs at 21 \AA\ and 12 \AA, respectively. 
Units were set to \AA, fs, and kcal/mol for length, time and energy respectively. All FES calculations were run for 50 ns.
In all cases, a 2 fs MD timestep was used.

\begin{table}
\caption{\label{tab:lj} Lennard-Jones parameters for cavity and ligand atoms. CW, CP, and CF refer to the anchor, cavity, and fullerene atoms respectively. The parameters for inter-molecular interactions  are described by combination rules: $\sigma_{ij} = \sqrt{\sigma_i \sigma_j}$ and $\epsilon_{ij} = \sqrt{\epsilon_i \epsilon_j}$}
\begin{ruledtabular}
\begin{tabular}{lccr}
i&j&$\sigma ({\rm nm})$&$\epsilon (\frac{{\rm kJ}}{{\rm mol}})$\\
\hline
CF & CF & 0.35 & 0.276144\\
CW & CW & 0.4152 & 0.00240\\
CP & CP & 0.4152 & 0.00800\\
CF & CW & 0.3812 & 0.02574\\
CF & CP & 0.3812 & 0.04700\\
\end{tabular}
\end{ruledtabular}
\end{table}

\subsubsection{Streptavidin-Biotin Complex}
\label{sec:details_bsa}
In Ref.~\onlinecite{tiwarybio} the dimeric version of the biotin-streptavidin complex was studied to determine an unbinding CV and compute an unbinding rate. Here we used the same system and calculated unbinding rates at several pulling forces.  For the SA/b atomistic system, a bound structure of biotin and a dimeric form of streptavidin was obtained from the protein data bank with PDB ID: 3RY2 \cite{strep} (Fig. \ref{fig:fig:bsaimg}). The all atom AMBER ff99SB*-ILDN \cite{Lindorff-Larsen2010} force field was used to describe all bonded and non-bonded interactions in the protein and the TIP4P model was used for water. The charged biotin ligand was parameterized with AM1-bcc charges and GAFF \cite{wang2004development} parameters as in Ref.~\onlinecite{tiwarybio}. 

The ligand and protein structures were combined and neutralized with counter ions. The complex was then solvated with TIP4P water and an ion concentration of 150 mM NaCl was added to the system to approximate physiological/experimental conditions. The full system’s energy was later minimized and subsequent NVT and NPT 1 ns equilibration was performed while restraining the complex in its bound pose.
The Nose-Hoover thermostat and the Parrinello-Rahman barostat were used in the NPT production runs at 300 K.

For rate calculations, an optimized one dimensional reaction coordinate reported in Ref.~\onlinecite{tiwarybio} was used as the collective variable for InfrMetaD. The reaction coordinate is a linear combination of two distances, $\chi = \psi_1 + 0.75\psi_2$. Where $\psi_1$ is the distance between the COM of the oxygen (OG) in residue S45 and the nitrogen in residue N49, and the COM of the C11 and N2 atoms in biotin, and $\psi_2$ is the distance between the carbon atom (CG) atom in residue D128 and the N1 atom in biotin. The unbound state was located at $\chi = 30 $ \AA. 

InfrMetaD was performed using HEIGHT=0.478, SIGMA=0.2 BIASFACTOR=15 and PACE=7500. The ligand was pulled along the distance between the COM of the binding pocket and the COM of the ligand with constant force. A total of 20 runs were performed for each force in the 0 to 72 pN range in intervals of 3 pN. 

FES estimates were obtained by running WTMetaD simulations for each force with HEIGHT=0.478, SIGMA=0.2 BIASFACTOR=12 and PACE=600. The bias was applied to the optimized reaction coordinate while the pulling force was applied to the 3D distance between the COMs of biotin and binding pocket (binding pocket consists of residues:L25, S27, Y43, S45, V47, G48, A50, W79, R84, A86, S88, T90, W92, W108, L110, and D128 following numbering in 3RY2\cite{strep} as in Ref.~\onlinecite{Rico6594}) . PLUMED walls were applied to both $\psi_1$ and $\psi_2$ at 22 \AA\  and 14 \AA\  respectively.

The FES estimate for the system at 0 force was obtained by running 20 such WTmetaD simulations for 100 ns each. The bias CVs were then reweighted and the FE was obtained via $F(Q)=-k_b T \ln(P(\vec{Q}))$ where P(\vec{Q}) is a weighted histogram of all \vec{Q} sampled in all the simulations combined. The same procedure was followed for the rest of the forces although only $13 \times 50$ ns simulations were performed for each. Units were set to \AA, fs, and kcal/mol for length, time and energy respectively. All simulations for this system were also performed in GROMACS \cite{gromacs} using a 2 fs MD timestep.

\begin{acknowledgments}
We thank Jagannath Mondal for providing GROMACS input files for the cavity-ligand model.
Computations were performed on resources provided via the NYU High Performance Computing center. G.M.H. was supported by the National Institutes of Health (NIH) via the award R35-GM138312. W.J.P.C. was initially supported by the Department of Energy via the award DE-SC0019695, and is now supported by NIH award R35-GM138312-02S1.

\end{acknowledgments}

\section*{Data Availability Statement}
Scripts and data to produce all figures in the manuscript, as well as input files to generate the data can be found at \url{https://github.com/hocky-research-group/PenaUnbindingPaper}

\bibliography{unbinding}

\renewcommand{\thefigure}{S\arabic{figure}}
\setcounter{figure}{0}

\clearpage
\onecolumngrid
\section{Supporting information}
\begin{figure*}[ht]
\centering
\includegraphics[width=0.8\columnwidth]{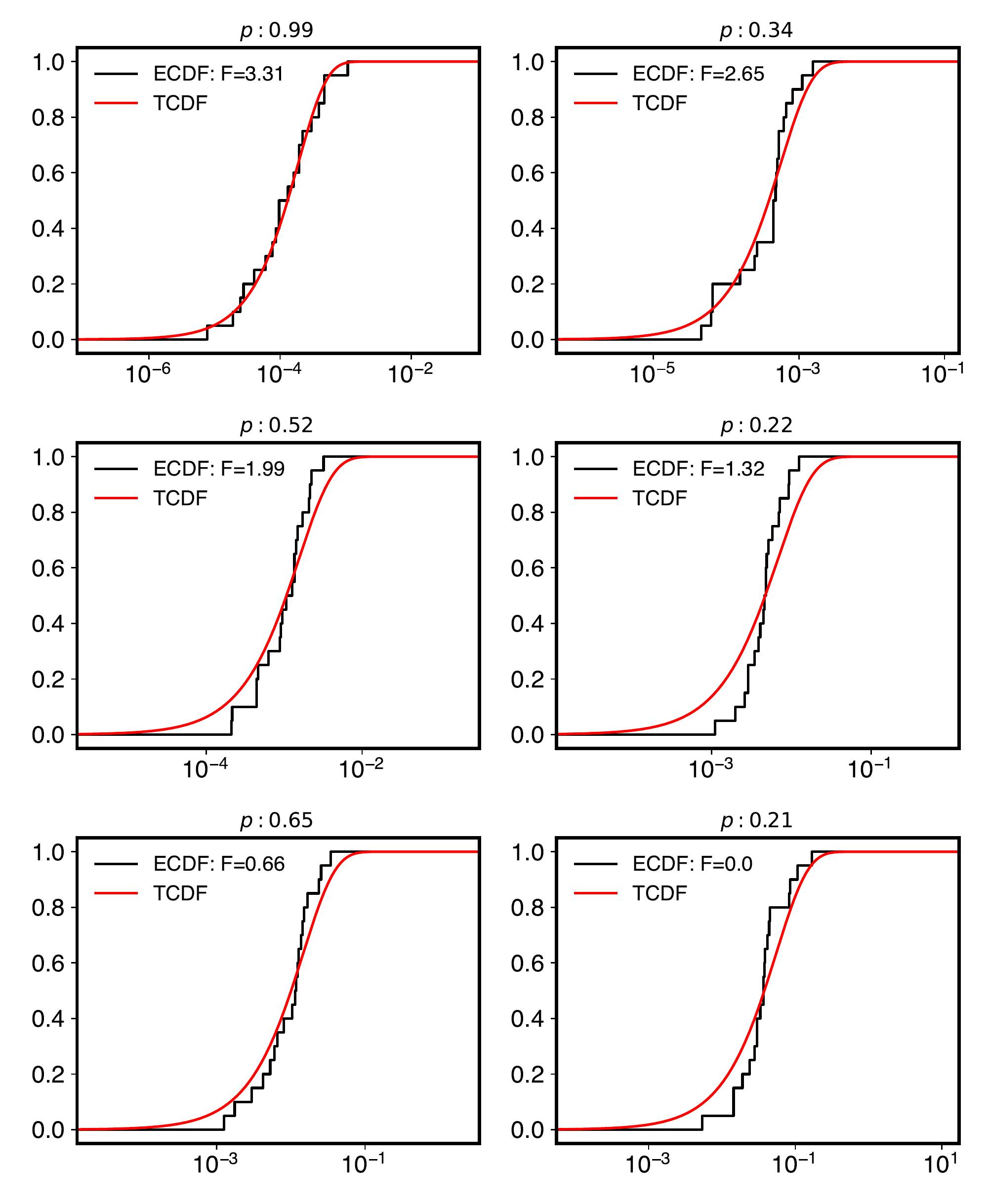}
\caption{\label{fig:cdfslip} Representative CDF fits are shown at various forces. For the simple 1D model the CDF fits are excellent and all fits for all forces pass the two-sample KS test.}
\end{figure*}

\begin{figure*}[ht]
\centering
\includegraphics[width=0.8\columnwidth]{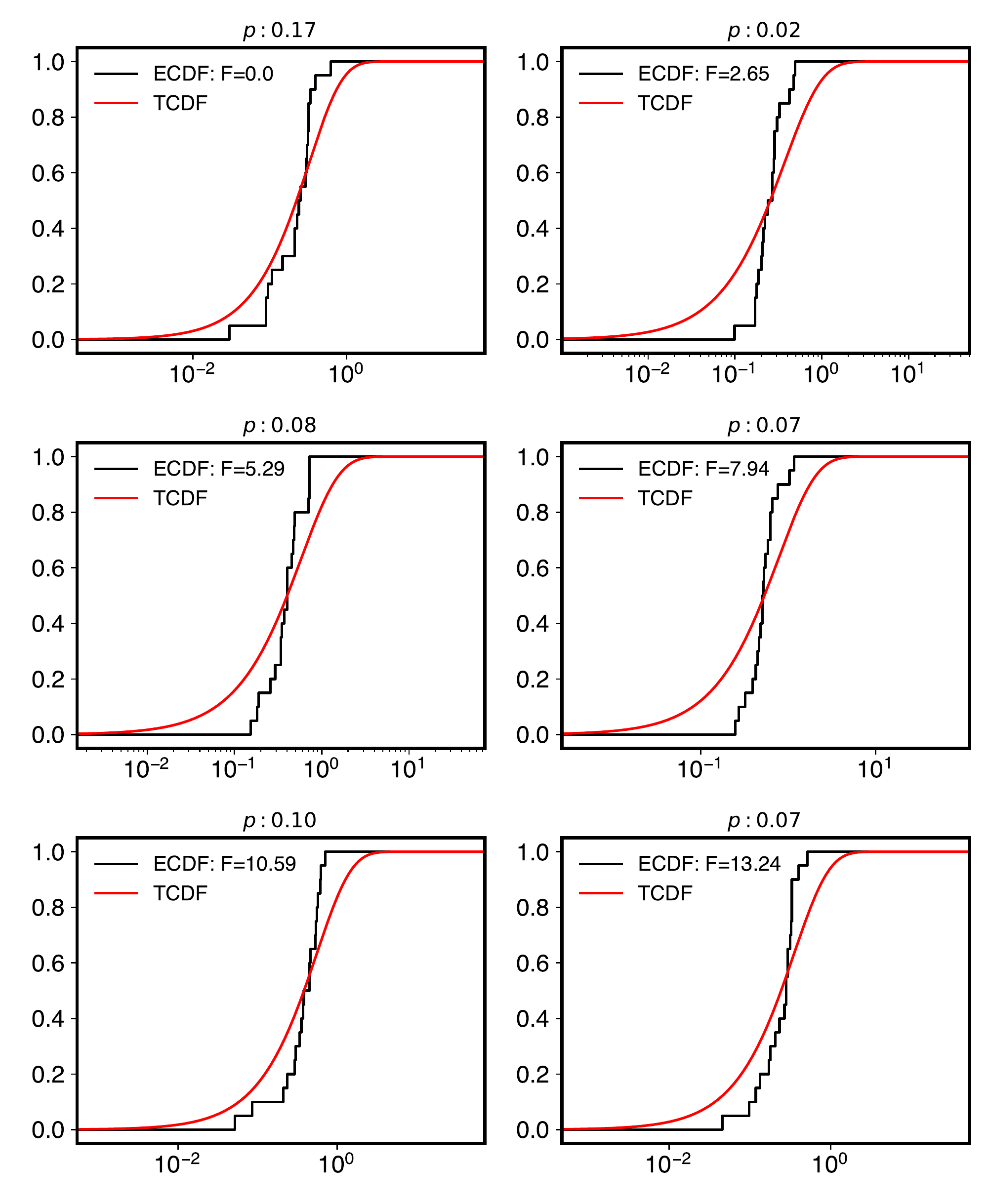}
\caption{\label{fig:cdfcatch}Representative CDFs fits are shown at various forces for the three-well system. Four out of 21 fits did not pass the two-sample KS tests; the rest of the CDF fits have $p>0.05$ and pass the two-sample KS test.}
\end{figure*}

\begin{figure*}[ht]
\centering
\includegraphics[width=0.8\columnwidth]{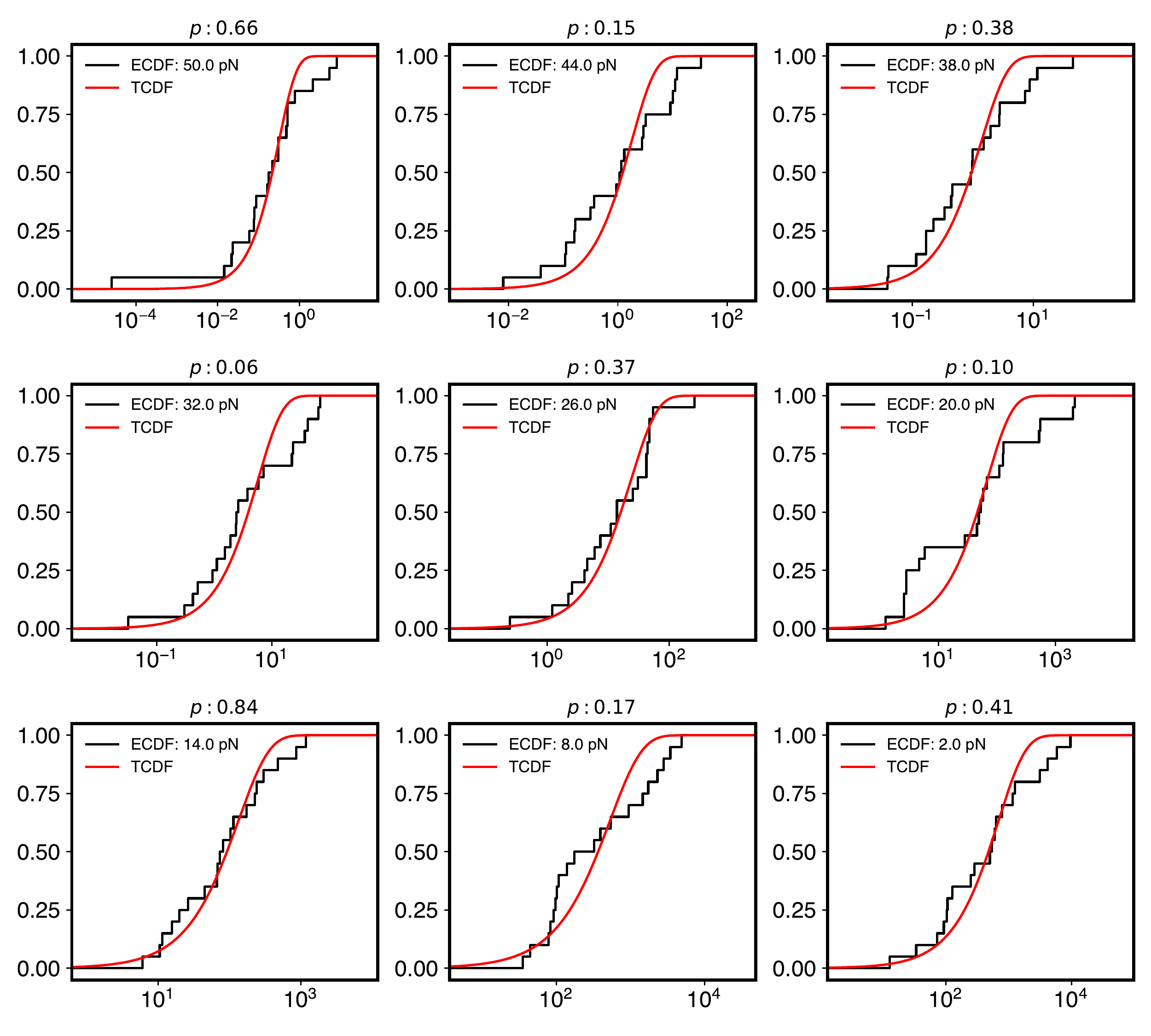}
\caption{\label{fig:cdfcavity} Representative CDF fits are shown at various forces for the cavity ligand system. One out of 26 fits did not pass the two-sample KS tests; the rest of the CDF fits are excellent and pass the two-sample KS test.}
\end{figure*}

\begin{figure*}[ht]
\centering
\includegraphics[width=0.8\columnwidth]{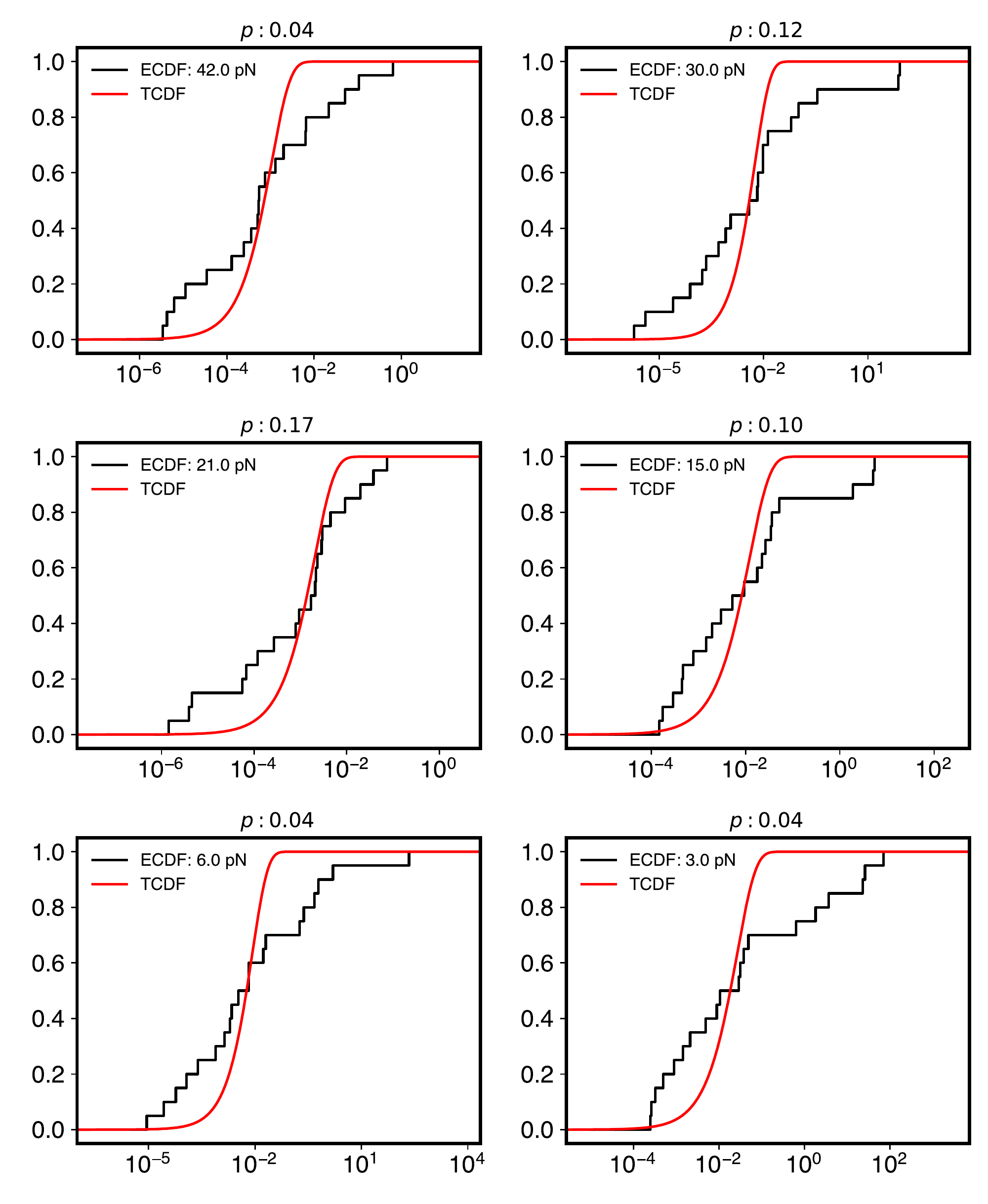}
\caption{\label{fig:cdfbsa} Representative CDFs fits are shown for the forces at which the CDF fit does and does not pass the two-sample KS test. Besides the three forces shown with $p>0.05$, all other fits failed the test.}
\end{figure*}

\begin{figure*}[ht]
\centering
\includegraphics[width=0.8\columnwidth]{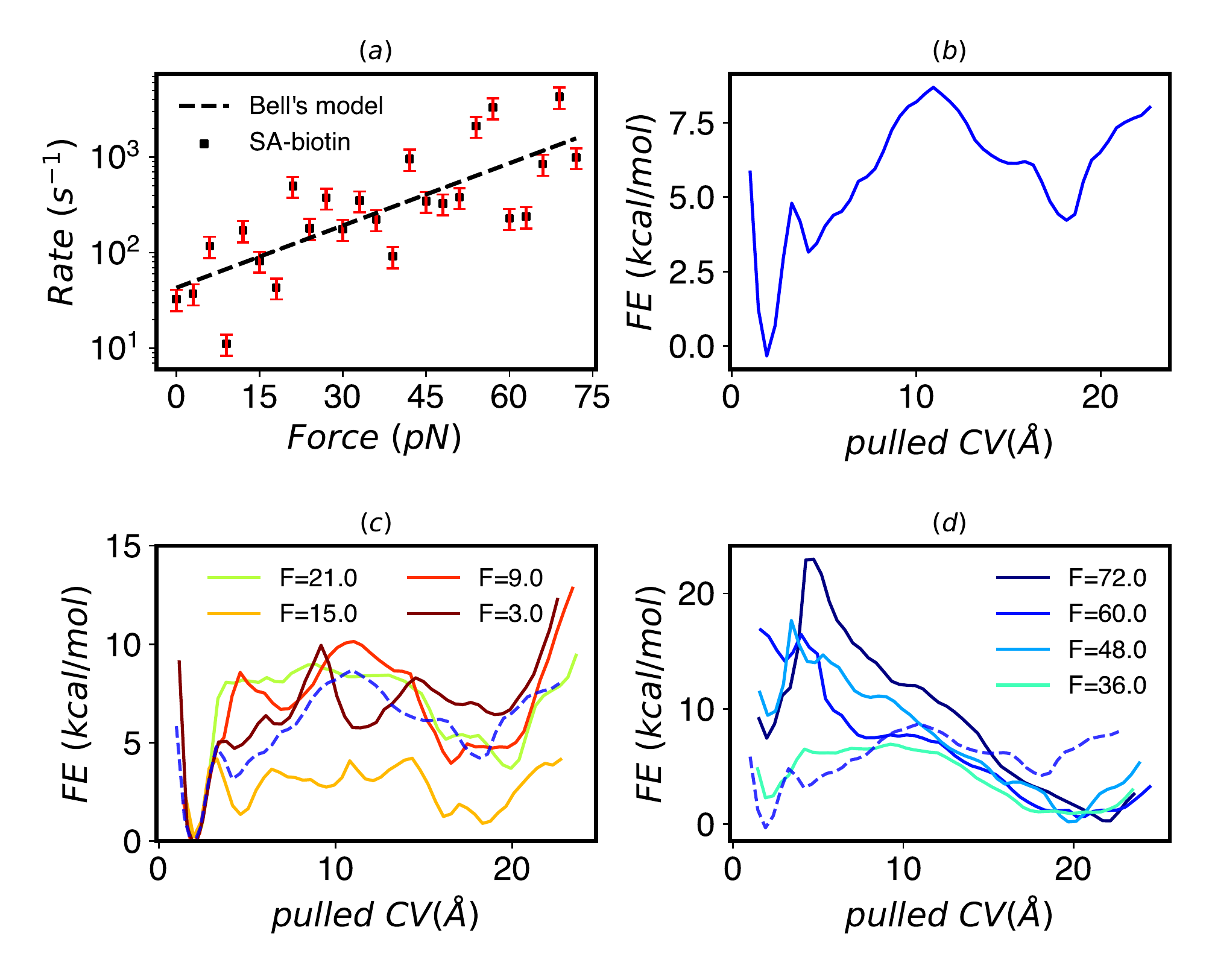}
\caption{\label{fig:fesbsapull} FES estimate as a function of pulling CV. This CV is the 3D distance between the binding pocket and biotin. Similarly to what is seen in Fig.~\ref{fig:bsalograte}, the unbound state is far away from the bound state and  more apparent intermediates appear. At larger forces the free energy favors the unbound states and the intermediate states are not apparent.}
\end{figure*}
\end{document}